\documentclass[]{aastex631}
\usepackage{graphicx}
\usepackage{color}
\usepackage{epsfig}
\usepackage{subfigure}

\usepackage{amssymb}
\usepackage{amsmath}
\usepackage{natbib}
\usepackage{graphicx}
\usepackage{color}
\usepackage{tabularx}
\usepackage{epsfig}
\usepackage{rotating}
\usepackage{multirow}
\usepackage{subfigure}
\usepackage{ulem}
\usepackage{longtable}

\shorttitle{Galacitc disk from Cepheid kinematics}
\shortauthors{X. Zhou, X. Chen, L. Deng, S. Wang}
\graphicspath{ {Figures/} }

\begin{document}

\title{Tracing the Galactic disk from the kinematics of Gaia Cepheids}

\correspondingauthor{Xiaoyue Zhou, Xiaodian Chen}
\email{zhouxiaoyue22@mails.ucas.ac.cn, chenxiaodian@nao.cas.cn}

\author{Xiaoyue Zhou}
\affiliation{CAS Key Laboratory of Optical Astronomy, National Astronomical Observatories, Chinese Academy of Sciences, Beijing 100101, China}
\affiliation{School of Astronomy and Space Science, University of the Chinese Academy of Sciences, Beijing, 100049, China}

\author{Xiaodian Chen}
\affiliation{CAS Key Laboratory of Optical Astronomy, National Astronomical Observatories, Chinese Academy of Sciences, Beijing 100101, China}
\affiliation{School of Astronomy and Space Science, University of the Chinese Academy of Sciences, Beijing, 100049, China}
\affiliation{Department of Astronomy, China West Normal University, Nanchong, 637009, China}
\affiliation{Institute for Frontiers in Astronomy and Astrophysics, Beijing Normal University, Beijing 102206, China}

\author{Licai Deng}
\affiliation{CAS Key Laboratory of Optical Astronomy, National Astronomical Observatories, Chinese Academy of Sciences, Beijing 100101, China}
\affiliation{School of Astronomy and Space Science, University of the Chinese Academy of Sciences, Beijing, 100049, China}
\affiliation{Department of Astronomy, China West Normal University, Nanchong, 637009, China}

\author{Shu Wang}
\affiliation{CAS Key Laboratory of Optical Astronomy, National Astronomical Observatories, Chinese Academy of Sciences, Beijing 100101, China}
\affiliation{Department of Astronomy, China West Normal University, Nanchong, 637009, China}

\begin{abstract}
Classical Cepheids (CCs) are excellent tracers for understanding the structure of the Milky Way disk. The latest Gaia Data Release 3 provides a large number of line-of-sight velocity information for Galactic CCs, offering an opportunity for studying the kinematics of the Milky Way. We determine the three-dimensional velocities of 2057 CCs relative to the Galactic center. From the projections of the 3D velocities onto the XY plane of the Galactic disk, we find that $V_{R}$ and $V_{\phi}$ velocities of the northern and southern warp (directions with highest amplitude) are different. This phenomenon may be related to the warp precession or the asymmetry of the warp structure. By investigating the kinematic warp model, we find that the vertical velocity of CCs is more suitable for constraining the warp precession rate than the line of nodes angles. Our results suggest that CCs at $12-14$ kpc are the best sample for determining the Galactic warp precession rate. Based on the spatial structure parameters of Cepheid warp from \cite{2019NatAs...3..320C}, we determine a warp precession rate of $\omega = 4.9\pm1.6$ km s$^{-1}$ kpc$^{-1}$ at 13 kpc, which supports a low precession rate in the warp model. In the future, more kinematic information on CCs will help to better constrain the structure and evolution of the Milky Way.
\end{abstract}


\section{Introduction} \label{sec:intro}
Exploring the structure and evolution history of the Milky Way is one of the most important topics in astronomy. Constructing geometric and kinematic models of the Milky Way is limited by our location in the Galaxy and the effects of interstellar extinction, so a complete picture cannot be constructed as easily as for other galaxies. The research on the structure of the Galactic disk has been broadly divided into radial and vertical directions. In the radial direction, since the 1950s, we have known that the Milky Way has a spiral arm structure through the distribution of local OB associations \citep{1953ApJ...118..318M}. Since then, many works have been devoted to studying the spiral arms of the Milky Way's disk \citep{2014ApJ...783..130R, 2016ApJ...823...77R, 2016SciA....2E0878X}. \cite{2019ApJ...885..131R} updated the logarithmic model of the spiral arms, and identified the main spiral arms as Norma-Outer arm, Sagittarius-Carina arm, Scutum-Centaurus arm, and Perseus arm, as well as an isolated segment, the Local arm. In addition, the Galactic disk has  strong asymmetry and rich structures in velocity space. In particular, the structure of arches and ridges can be seen in Galactocentric azimuthal velocity and radius space \citep{2018Natur.561..360A, 2018A&A...619A..72R, 2019MNRAS.488.3324F, 2019MNRAS.489.4962K, 2020ApJ...902...70W}. However, the dynamical nature of these phenomena and their formation mechanism remain unclear. Existing explanations include resonances with the bar \citep{2019MNRAS.484.3291T, 2019A&A...626A..41M, 2020A&A...643L...3L, 2021MNRAS.500.2645T} and/or processes with spiral arms \citep{2019A&A...626A..41M, 2020ApJ...888...75B} and/or external perturbations such as the Sagittarius dwarf galaxy perturbation \citep{2018MNRAS.481.1501B, 2019MNRAS.486.1167B, 2019MNRAS.489.4962K}.

Looking at the vertical structure of the Galactic disk, the observations of 21-cm neutral hydrogen \citep{1957AJ.....62...93K, 2006ApJ...643..881L, 2006astro.ph..1653V}, dust \citep{1994ApJ...429L..69F}, stars \citep{1990A&A...230...21W}, and stellar kinematics \citep{2018MNRAS.481L..21P, 2019A&A...627A.150R,2020ApJ...905...49C} all suggest that the Galactic disk is warped and flared. At present, the mechanism that causes the warp's formation is not well understood, and the relevant theoretical explanations can be divided into two categories. One is that the warp is the result of gravitational interactions \citep{2000A&A...354...67D}, including interactions with other satellite galaxies \citep{Kim_2014, 2019MNRAS.485.3134L} and a misaligned dark matter halo \citep{2014MNRAS.440.1971W, 2017A&A...602A..67A}. The other is non-gravitational mechanisms, such as the accretion of interstellar matter \citep{2002A&A...386..169L, 2020ApJ...897..119W} or interactions with intergalactic magnetic fields \citep{1990A&A...236....1B, 2010A&A...519A..53G}. Knowledge of stellar kinematics helps to study the origin of Galactic warp, such as the interaction with satellite galaxies \citep{2020NatAs...4..590P}, or torque of a misaligned non-spherical halo \citep{1999MNRAS.303L...7J}, or torque of the inner disk \citep{2019NatAs...3..320C}. The Galactic warp precession rate was measured for the first time using 12 million giants from Gaia Data Release 2 (DR2) by \cite{2020NatAs...4..590P}. They gave a value of precession $\beta=10.86\pm0.03$ (stat.) $\pm3.20$ (syst.) km s$^{-1}$ kpc$^{-1}$, and supported the scenario that the warp is the result of ongoing encounters with satellite galaxies. A counter-example is given by \cite{2021ApJ...912..130C}, who obtained the precession rate $\beta=4\pm5$ km s$^{-1}$ kpc$^{-1}$, indicating that the warp precession rate is not that high. They argued that there is a systematic error in the precession rate calculated by \cite{2020NatAs...4..590P}. The precession of Milky Way's warp is still an unsolved problem.

Classical Cepheids (CCs) are a class of pulsating variable stars whose luminosity varies linearly with the period. This relationship is called the period--luminosity relation (PLR), which was first discovered by \cite{1912HarCi.173....1L}. CC's absolute magnitude can be determined by the PLR. Combining the observed apparent magnitude, the distance can be determined. Therefore, CC is the standard candle and its PLR is an essential component of the cosmic distance ladder for precise measurement of the Hubble constant \citep{2012ApJ...758...24F, 2022ApJ...938...36R}. Meanwhile, CCs are young stars found mainly in the spiral arms, the disk and open clusters of the Milky Way. CCs can be used to trace structures of the disk and spiral arms. The age information of CCs can be obtained from the CC's period--age relation \citep{2021MNRAS.504.4768Z}. Therefore, CCs are an important tool for modeling the structure and evolution history of the Galactic disk \citep{2013A&A...558A..31L, 2019NatAs...3..320C, 2019Sci...365..478S, 2019AcA....69..305S, 2022A&A...668A..40L}. In the past five years, \cite{2019NatAs...3..320C} and \cite{2019Sci...365..478S} demonstrated the intuitive three-dimensional (3D) structure of the Galactic disk for the first time based on CCs. Among them, \cite{2019NatAs...3..320C} discovered the line of nodes (LONs) of warp do not always lie in the same direction, showing a pattern of leading spiral. \cite{2019Sci...365..478S} revealed the evolution of spiral arms. Besides, \cite{2021A&A...654A.138M} used CCs to extend to the far side disk of the Milky Way, while \cite{2022A&A...668A..40L} used CCs to investigate the over-density regions in spiral arms.

Nowadays, the Gaia DR3 provides the mean line-of-sight velocity (RV) information for the Milky Way's CCs, creating new opportunities to understand the Milky Way \citep{2023A&A...674A..37G}. A new resonance-like feature was discovered based on data from the Gaia DR3 CCs \citep{2023A&A...670A..10D, 2023MNRAS.519..902S}. Therefore, the goal of this paper is to study the Galactic outer disk based on the kinematic information of Gaia CCs, including the rotation curve, the Galactic warp and flare. In this paper, we present the data and data processing procedures including velocity calculation and criteria in Section \ref{sec:data}. The analysis of the asymmetric structure of the Milky Way and the warp precession model is discussed in Section \ref{sec:result}. The summary of this work is in Section \ref{sec:concl}.

\section{Data} \label{sec:data}

In this section, we focus on the processing of the original data from Gaia DR3 to obtain the 3D position and velocity parameters of the CCs in the Galactic cylindrical coordinate. Based on velocity histograms and error analysis, we exclude possible outliers in the velocities and obtain the final sample for subsequent analysis.
\subsection{Raw sample}
\label{sec:2.1}

The European Space Agency’s (ESA) Gaia mission released the Gaia DR3 catalog in June 2022. The celestial position, proper motion, parallax, and average photometry in the $G$, $G_{Bp}$, and $G_{Rp}$ bands of the source were given in Gaia EDR3. In contrast, Gaia DR3 included newly determined mean RV for more than 33 million stars, atmospheric parameters and chemical abundance for about 32.2 million stars, and provided multi-band time series photometry for nearly 12 million variable stars \citep{2023A&A...674A...1G, 2023A&A...674A..13E}. The gaiadr3.vari\_cepheid catalog given by Gaia DR3 contains 15021 different types of CCs, which are distributed in different galaxies such as Large Magellanic Cloud (LMC), Small Magellanic Cloud (SMC), Milky Way, M31, M33, etc. We obtained these CCs' astrometric parameters by cross-matching the gaiadr3.vari\_cepheid catalog with the gaiadr3.gaia\_source catalog from the Gaia archive. \cite{2023A&A...674A..17R} pointed out that there is a sub-sample of 799 CCs of all types has time-series RV curves, and the typical uncertainties in $\langle RV \rangle$ are about $1-1.5$ km s$^{-1}$. To get a larger sample, \cite{2023A&A...674A..37G} found that the average RV values of CCs calculated by the spectroscopic pipeline in the gaia\_source catalog are in overall agreement with the RV curve fitting values in the vari\_cepheid catalog, with a mean difference of $0.6\pm6.0$ km s$^{-1}$. They provided a catalog of a total of 3306 Galactic CCs, including a small number of CCs collected from the literature. We then cross-matched the previously obtained table of CCs with these 3306 CCs and obtained a total of 3038 CCs, of which 2057 have RV information. These subsamples with RVs serve as our initial Gaia sample, which includes 1364 fundamental CCs, and 693 first-overtone CCs. The metallicities and errors for all CCs are from \cite{2023A&A...674A..37G}, which were estimated by the radial gradient of the Galactic disk metallicity determined by CCs: ${\rm [Fe/H]}=(-0.0527 \pm 0.0022)R+(0.511 \pm 0.022)$, ${\rm rms}=0.11\ {\rm dex}$ \citep{2022A&A...659A.167R}. We now have a catalog including the spatial coordinates, proper motions, mean RVs, metallicities, periods, and mean intensity magnitude in three bands for 2057 CCs.

\subsection{Calculation of the distance and velocity of CCs}
\label{sec:2.2}

The Gaia parallax has a zero-point offset, even after correction \citep{2021A&A...649A...4L}. Many studies have been devoted to the study of the zero-point deviation of the Gaia parallax \citep{2021ApJ...911L..20R, 2021A&A...654A..20G, 2021AJ....161..214Z}. Gaia parallaxes can achieve high accuracy within $1-2$ kpc, and the parallax error is $10\%-20\%$ for stars within 4 kpc. But relative errors of parallaxes are larger at greater distances, which may affect the study of the outer disk of the Milky Way. Meanwhile, the improved accuracy of Gaia photometry (error at the level of 1-10 mmag) and the Gaia-band period–Wesenheit relation of the CCs provide opportunities for us to estimate more accurate distances of CCs. Therefore, we used the PLR-based distance instead of the Gaia parallax. We adopted the mean intensity magnitude of three bands ($\langle G \rangle, \langle Bp \rangle, \langle Rp \rangle$). According to $W_{G,Bp, Rp}= \langle G \rangle - 1.90(\langle Bp \rangle - \langle Rp \rangle)$ given by \cite{2019A&A...625A..14R}, the Wesenheit magnitude of the Gaia bands was obtained, which reduces the effect of extinction. The Gaia-band absolute magnitude was obtained by the period–Wesenheit-metallicity relation \citep{2022A&A...659A.167R}:

   \begin{equation} \label{eq1}
      \langle W_{G,Bp, Rp} \rangle = (-5.988 \pm 0.018) - (3.176 \pm 0.044)(\log P - 1.0) - (0.520 \pm 0.090){\rm [Fe/H]}     \\
   \end{equation}
   
For the first-overtone CCs, we converted their first-overtone periods ($P_{1O}$) to the corresponding fundamental periods ($P_{\rm F}$) through the equation of \cite{1997MNRAS.286L...1F}: $P_{\rm F} = P_{\rm 1O}/(0.716 - 0.027\log P_{\rm 1O})$, and then estimated their absolute Wesenheit magnitudes using Equation (\ref{eq1}). Combined with the mean intensity magnitude of three bands, the distance modulus of each CC is estimated by $\mu_{0} = m_{W} - M_{W}$. Then we used the following formula to convert the distance modulus to parallax in units of mas and distance in units of kpc: 

   \begin{equation} \label{eq2}
   \begin{aligned}
      &\varpi_{{\rm phot},i}\ ({\rm mas}) = 10^{-0.2(\mu_{0}-10)}; \\
      &\textsl{d}\ ({\rm kpc}) = 1/\varpi_{{\rm phot},i} \\
   \end{aligned}
   \end{equation}

Now we have the distance $d$ from each CC to the sun. Combining the Galactic longitude and latitude ($l, b$), we projected the heliocentric distances $d$ into the 3D Cartesian coordinate ($x, y, z$) centered on the sun by the equation: 

   \begin{equation} \label{eq3}
     \begin{pmatrix}
     x \\
     y \\
     z \\
     \end{pmatrix}
     = d\begin{pmatrix}
     \cos b \cos l \\
     \cos b \sin l \\
     \sin b \\
     \end{pmatrix}
   \end{equation}

Here, the positive $x$ direction is towards the center of the Milky Way. Then we assumed the sun is 8 kpc away from the Galactic center in order to use the warp model parameters of \cite{2019NatAs...3..320C}, and located in the mid-plane ($Z = 0$) to obtain the Galactocentric Cartesian coordinates ($X, Y, Z$) through coordinate translation, where $X = -x+8$. In this way, we calculated the Galactocentric distance $R$ and azimuth $\phi$ of each CC in the Galactic cylindrical polar coordinates through $R = \sqrt{X^{2}+Y^{2}}$ and $\phi = \arctan(\frac{Y}{X})$, and $\phi$ increases with the rotation direction of the Milky Way. To verify the validity of the distances, we converted the heliocentric distances from \cite{2023A&A...674A..37G} to Galactocentric distances. By comparison, we found that 98\% of the stars have a distance deviation of less than 0.1 kpc.

To obtain 3D velocity components in the Galactic cylindrical polar coordinate, we extracted the sample's proper motion and mean RV data ($\mu_{\alpha^{*}}, \mu_{\delta}, v_{r}$) , where $\mu_{\alpha^{*}}$ = $\mu_{\alpha}\cos\delta$. We converted the ($\mu_{\alpha^{*}}$, $\mu_{\delta}$) in mas yr$^{-1}$ to the tangential velocity in km s$^{-1}$ by multiplying the stellar heliocentric distance $d$ and 4.74047. We multiplied it by the matrix conversion factor $A'_{G}$ provided by the Gaia EDR3 online documentation to obtain the velocity in the Galactic coordinate. Then, we multiplied the matrix of the normal triad $A$:

   \begin{equation} \label{eq4}
     A = \begin{pmatrix}
     -\sin\alpha&-\sin\delta\cos\alpha&\cos\delta\cos\alpha \\
     \cos\alpha&-\sin\delta\sin\alpha&\cos\delta\sin\alpha \\
     0&\cos\delta&\sin\delta \\
     \end{pmatrix}
   \end{equation}

The velocity of the heliocentric Cartesian coordinates is estimated by:

   \begin{equation} \label{eq5}
     V_{rel} = \begin{pmatrix}
     u \\
     v \\
     w \\
     \end{pmatrix}
     = A'_{G}A\begin{pmatrix}
         4.74047 \mu_{\alpha^{*}}d \\
         4.74047 \mu_{\delta}d \\
         v_{r} \\
     \end{pmatrix} \\
   \end{equation}

The detailed velocity calculation equation, including the matrix information, was described in Section 3.2 of \cite{2023A&A...674A..37G}. Here, we adopted the solar motion velocity toward the Galactic center $[U_{\odot}, V_{\odot}, W_{\odot}] = [9.3 \pm 1.3,\ 251.5 \pm 1.0,\ 8.59 \pm 0.28]\ $ km s$^{-1}$, which is consistent with \cite{2023A&A...674A..37G} and \cite{2023A&A...670A..10D}, to convert $V_{rel}$ to a reference centered on the Milky Way. Therefore, we got the 3D velocity of the CC relative to the Milky Way by $[u_{*}, v_{*}, w_{*}] = V_{rel}+[U_{\odot}, V_{\odot}, W_{\odot}]$. The positive direction of $u_{*}$ here is towards the Galactic center.
To obtain the velocity component in the Galactic cylindrical polar coordinate, we decomposed the velocity in the Cartesian coordinate by 

   \begin{equation} \label{eq6}
   \begin{aligned}
     &V_{R} = -u_{*}\cos(\phi)+v_{*}\sin(\phi); \\
     &V_{\phi} = u_{*}\sin(\phi)+v_{*}\cos(\phi); \\
     &V_{Z} = w_{*} \\
   \end{aligned}
   \end{equation}

Based on the azimuth angle $\phi$, we calculated the 3D velocity in the cylindrical polar coordinate. $V_{R}$, $V_{\phi}$ and $V_{Z}$ in Equation (\ref{eq6}) are the radial velocity, the azimuthal velocity and the vertical velocity, respectively. We now have the geometric and kinematic parameters to study the Galactic disk.

\subsection{Excluding outliers}
\label{sec:2.3}

\begin{figure*}
\centering
  \begin{minipage}{185mm}
  \includegraphics[width=185mm]{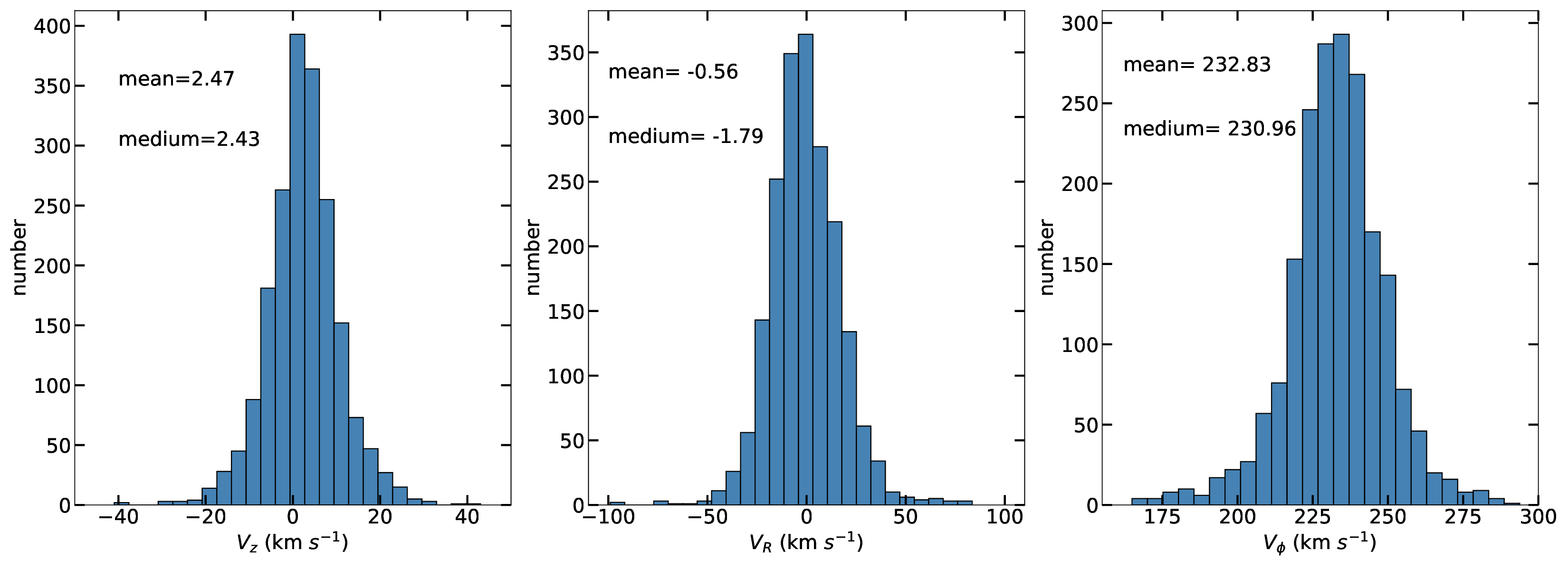}
\caption{3D velocity histogram of 1967 CCs in the Galactic cylindrical polar coordinate. The mean and median values of the velocities in each direction are labeled. \label{2.3.1.fig}}
  \end{minipage}
\end{figure*}

By visually inspecting the histogram of the velocity distributions of the 2057 Galactic CCs in the three directions, we found that some samples significantly deviate from the distribution of the overall sample. Therefore, we first excluded outliers using $V_{Z}$ in [$-50, 50$] km s$^{-1}$, $V_{R}$ in [$-100, 100$] km s$^{-1}$ and $V_{\phi}$ in [$160, 300$] km s$^{-1}$ as criteria. After selection, there are 1967 CCs left in the sample, including both fundamental and first-overtone CCs. The velocity distributions of the remaining CCs are shown in Fig. \ref{2.3.1.fig}. The mean and median values of velocities in different directions are also labeled in the figure. We found only 4 CCs with Galactocentric distances less than 4 kpc, and 6 CCs with Galactocentric distances greater than 19 kpc. The number is too small for us to study the nature of the Galactic disk there, so we excluded them, leaving 1957 CCs.

\begin{figure*}
  \begin{minipage}{185mm}
  \includegraphics[width=185mm]{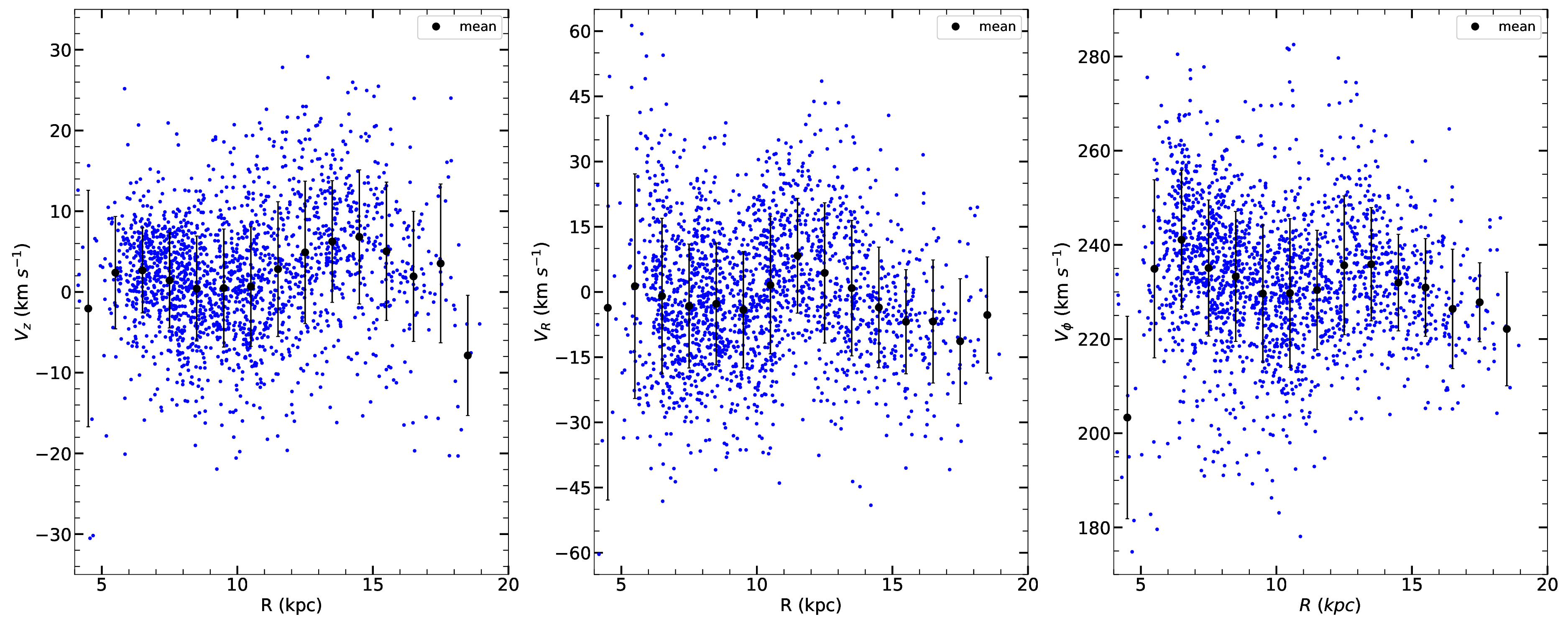}
\caption{3D velocity distribution of 1901 CCs at different Galactocentric distances. We used $3\sigma$ standard deviation as a criterion to exclude outliers. The black solid dots give the average value of each kpc. Error bars are standard deviations.  \label{2.3.2.fig}}
  \end{minipage}
\end{figure*}

After that, we calculated the average velocities of the CCs in each 1 kpc bin and plotted the variation of the 3D velocities with the Galactocentric distance as shown in Fig. \ref{2.3.2.fig}. It can be seen that some outliers deviate from the mean velocity, so we estimated the standard deviation of the velocity in each bin and excluded these outliers by the 3$\sigma$ clipping method. This procedure was performed for the velocities in the $V_{Z}$, $V_{R}$ and $V_{\phi}$ directions, and the remaining sample contained 1904 CCs. We noted that there are still very few stars with very large radial velocities at 4-6 kpc, so we excluded 3 CCs by the criterion of $V_{R}$ in [$-65, 65$] km s$^{-1}$. Fig. \ref{2.3.2.fig} illustrates the distribution of the 3D velocities of the final 1901 CCs at different Galactocentric distances. We can see that the blue dots are evenly distributed around the mean velocities (black filled circles). We will analyze the velocity features in the next section. 
 
\section{Analysis of the structure of the Galactic disk} \label{sec:result}

In this section we focus on the investigation of the different structures of the Milky Way using kinematic velocities of CCs. By analysing the projection of the distribution of the velocities on the Galactic plane and the variation of the average velocities along the Galactocentric distance, we discuss the structure of the north-south asymmetry of the Galactic disk. We also discuss the warp model of the outer disk.
\subsection{Velocity asymmetric structure}
\label{sec:3.1}

\begin{figure*}
\centering
  \begin{minipage}{135mm}
  \includegraphics[width=135mm]{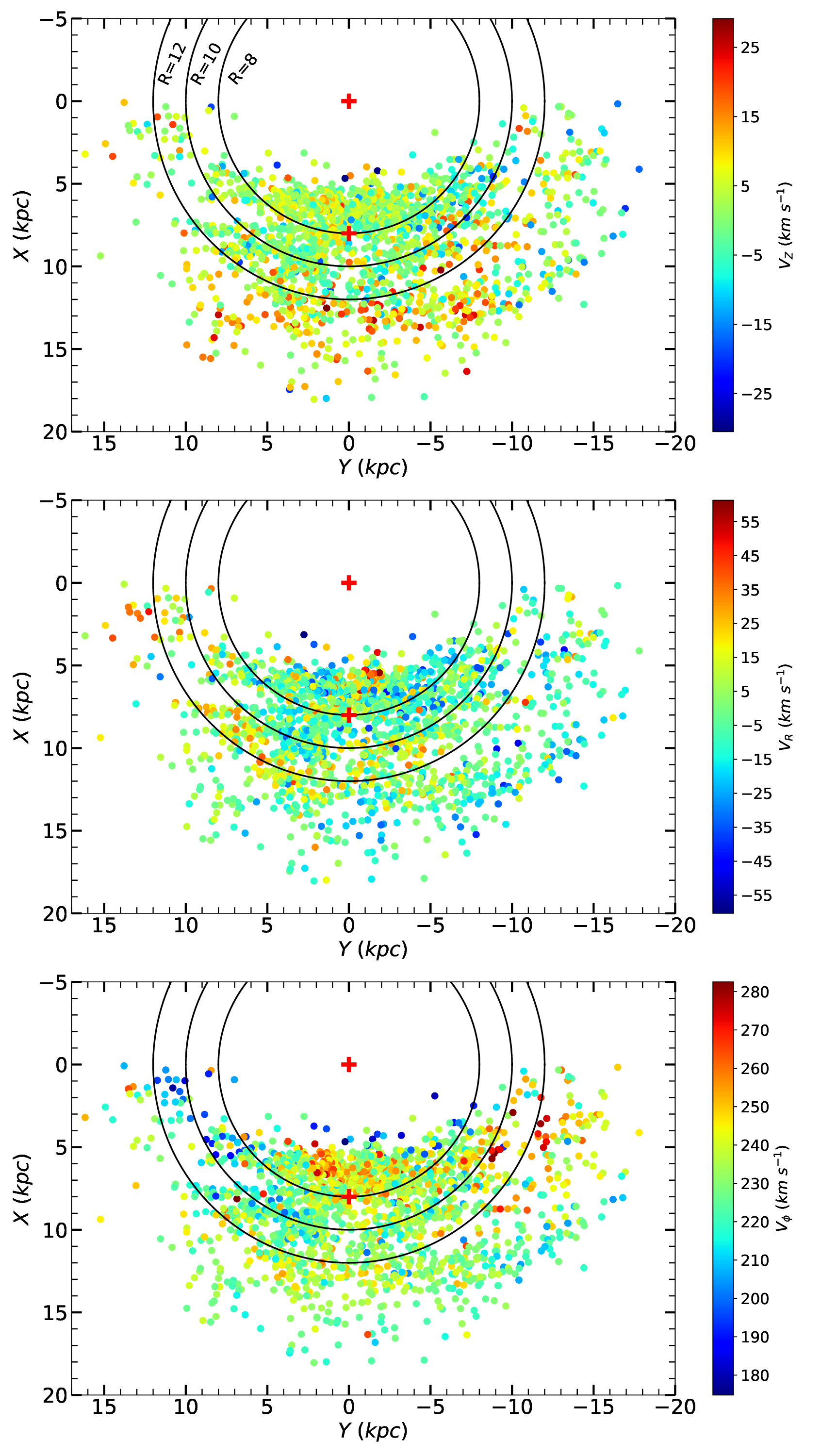}
\caption{Projection of CCs' 3D velocities in the $XY$ plane. From top to bottom, the colors are coded by $V_{Z}$, $V_{R}$, and $V_{\phi}$, where red represents larger velocities and blue represents smaller velocities. The position of the Galactic centre and the Sun are marked with red plus signs, and the solid black circles represent Galactocentric distances of 8, 10, and 12 kpc, respectively. \label{3.1.1.fig}}
  \end{minipage}
  \end{figure*}

CCs are important tracers for the structure of the Galactic disk due to their precise distances and wide distribution in the Milky Way. \cite{2019NatAs...3..320C} and \cite{2019AcA....69..305S} used the 3D spatial distribution of CCs to visualize the geometric structure of the Galactic disk. Based on the newly obtained RV and proper motion data, an updated kinematic map of CCs is available and provides new clues to the archaeology of the Galactic disk. The velocity projection in the $XY$ plane visualizes the asymmetry of the northern and southern disks, and the statistical effect of the 3D velocity at different Galactocentric distances reveals the structure of the Milky Way. 
In this section, we analyze the velocity projection in the $XY$ plane in Fig. \ref{3.1.1.fig} and the velocity variation along the Galactocentric distance in Fig. \ref{2.3.2.fig} to explore the distribution of 3D velocities and the asymmetric structure in the Galactic disk. 

In Fig. \ref{3.1.1.fig}, we project the velocity of 1901 CCs on the $XY$ plane of the Galactic disk, with the Y-positive direction being the northern disk and the Y-negative direction being the southern disk. Overall, the first thing we see is that the 3D velocities have a clear asymmetry in the northern and southern outer disks. This phenomenon is discussed in the last part of this subsection. From the projection of the vertical velocities on the Galactic disk in the top panel of Fig. \ref{3.1.1.fig} we see that $V_{Z}$ is dominated by positive velocities (solid circles in orange) outside of $R=12$ kpc, with a significant increase in velocity compared to the internal velocities.  This trend is more significant in the left panel of the Fig. \ref{2.3.2.fig}: the average value of $V_{Z}$ in the region from 5 kpc to 11 kpc is about 1.4 km s$^{-1}$, and then the average velocity gradually increases from 11 kpc to 14 kpc. The variation of $V_{Z}$ with $R$ in both figures indicates that the Milky Way is warped in the vertical direction, which is consistent with the warp in morphology. Meanwhile, in the left panel of Fig. \ref{2.3.2.fig}, we find that the vertical velocity dispersion gradually increases from $5.0-6.0$ km s$^{-1}$ to $8.0-9.0$ km s$^{-1}$ from the inner disk to the outer disk. This suggests that the Galactic disk thickens vertically with increasing Galactocentric distance, which is consistent with the flare feature of the Galactic disk, reflected in thin and thick disks whose scale height is larger in the outer regions \citep{2022A&A...664A..58C, 2019ApJ...871..208L, 2018MNRAS.478.3367W, 2014Natur.509..342F}.

The middle panel of Fig. \ref{3.1.1.fig} shows the projection of $V_{R}$ in the $XY$ plane. We see that radial velocities are dominated by positive velocities (solid circles in orange) at about $10-13$ kpc, indicating that there are more stars moving outward than inward in this region. This feature is more pronounced in the northern disk, indicating the asymmetry in the velocity structure between the northern and southern disks of the Milky Way. When we look at the middle panel of Fig. \ref{2.3.2.fig}, we see a peak shape at $10-14$ kpc, dominated by positive radial velocities, with more stars moving outward than inward in this region, and the opposite in other regions. This is consistent with Fig. \ref{3.1.1.fig}. Within $R<10$ kpc, the statistical average of the radial velocities is -1.9 km s$^{-1}$, when $R>10$ kpc, the average radial velocity of the stars starts to increase and reaches a maximum value of 8.3 km s$^{-1}$ near 12 kpc, after which the average radial velocity decelerates and reverses motion again. Within 15 kpc, the maximum average values of inward and outward velocities are 4.1 km s$^{-1}$ and 8.3 km s$^{-1}$, respectively. The overall characteristics of the radial velocity are in good agreement with \cite{2019AJ....157...26L, 2020MNRAS.491.2104W,2019A&A...621A..48L}.

In the right panel of Fig. \ref{2.3.2.fig}, we obtain the observed rotation curve of the outer Galactic disk. The results show that the rotation curve is generally flat with local fluctuations. The average rotational velocity decreases with a gradient of $-3.81$ km s$^{-1}$ kpc$^{-1}$ between 6 and 10 kpc, reaching a minimum of 229.66 km s$^{-1}$ at 9.5 kpc. Then the rotation curve rises with a gradient of about 1.56 km s$^{-1}$ kpc$^{-1}$ until it reaches a maximum value of 235.88 km s$^{-1}$ at 13.5 kpc, after which the rotational velocity gradually decreases. The shape of our rotation curve is consistent with \cite{2016MNRAS.463.2623H} and \cite{2021A&A...649A...8G}, where \cite{2016MNRAS.463.2623H} used the sample of red clump giants and K giants, and \cite{2021A&A...649A...8G} used extremely young massive stars with ages younger than 0.2 Gyr.

\begin{figure*}
\centering
  \begin{minipage}{185mm}
  \includegraphics[width=185mm]{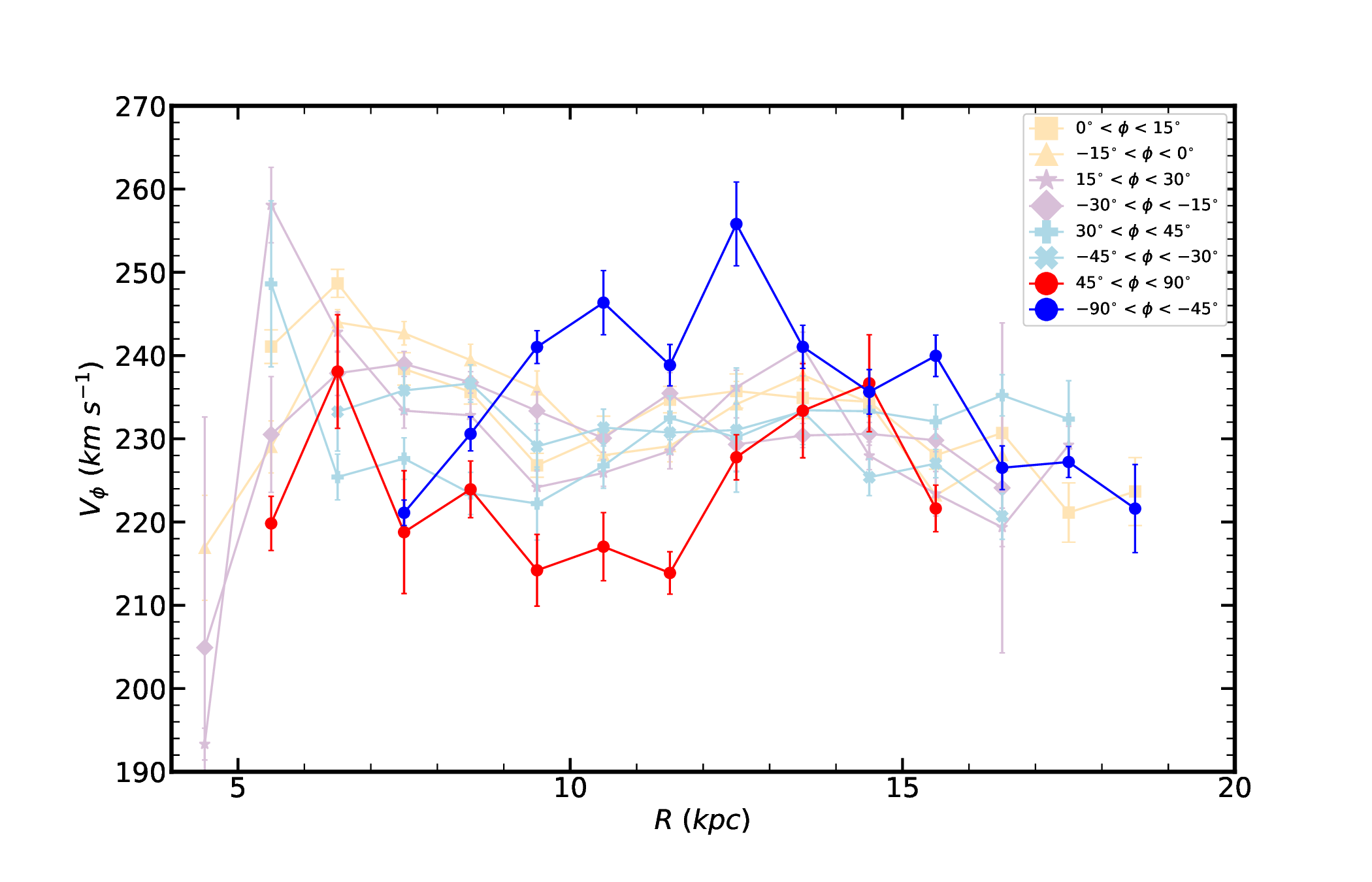}
  
\caption{Galactic rotation curves at different azimuth. We divided the azimuth into 8 ranges as shown in the legend. The northern and southern disk asymmetry is significant with azimuthal angles of $45^\circ<\phi<90^\circ$ and $-90^\circ<\phi<-45^\circ$. \label{3.1.2.fig}}
  \end{minipage}
\end{figure*}

\begin{figure*}
\centering
  \begin{minipage}{185mm}
  \includegraphics[width=185mm]{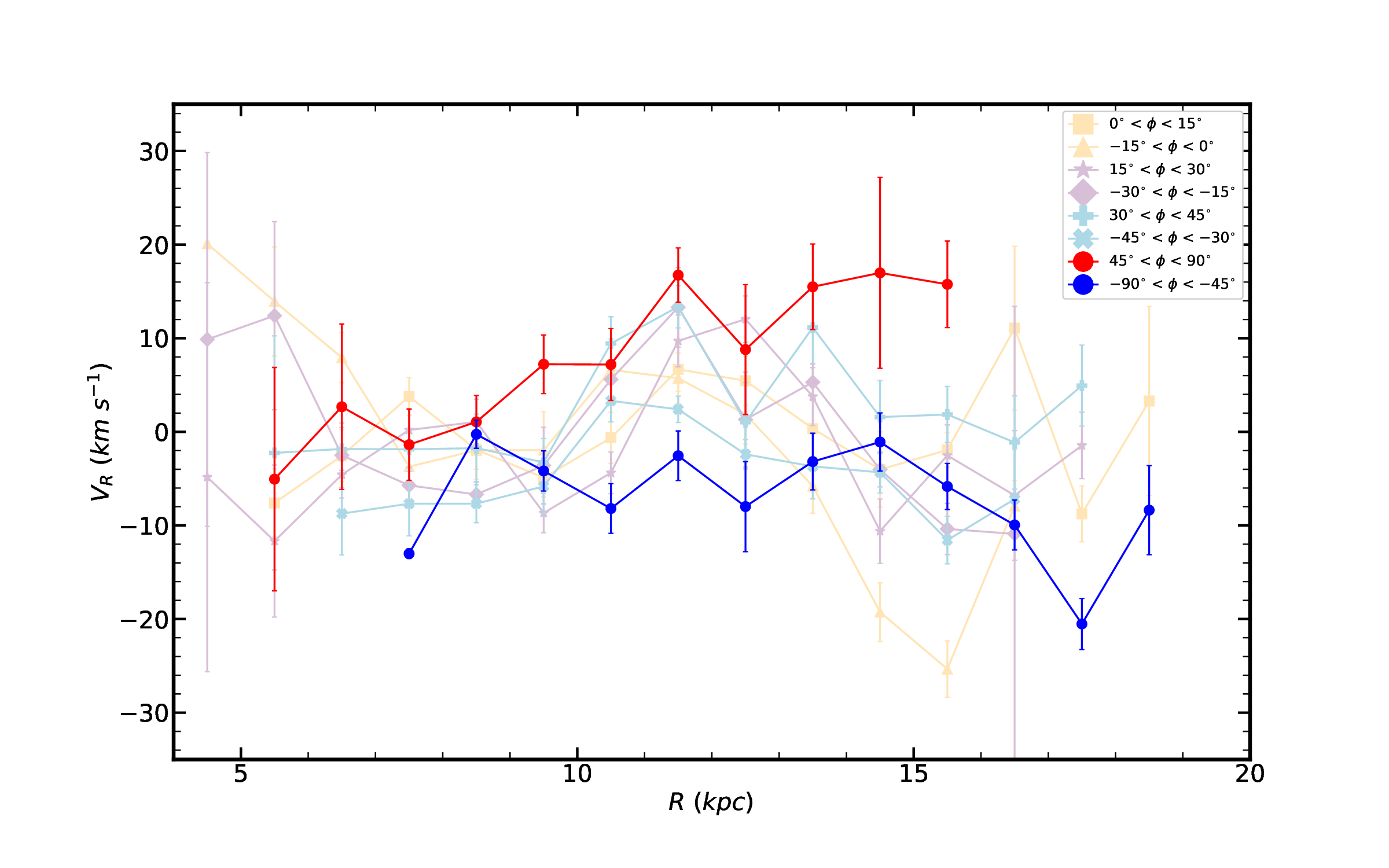}
  
\caption{Similar to Fig. \ref{3.1.2.fig} but for radial velocities. \label{3.1.3.fig}}
  \end{minipage}
\end{figure*}

In the bottom panel of Fig. \ref{3.1.1.fig}, rotational velocities present a clear asymmetry on the northern and southern warp in the directions with azimuthal angles around $90^\circ$ and $-90^\circ$. Here, we define the northern and southern warps to directions around $90^\circ$ and $-90^\circ$. We analyse the rotation curves for eight different azimuthal angle ranges in Fig. \ref{3.1.2.fig}. In each range, the sample size is similar, and we calculated the average rotational velocity within each 1 kpc bin of Galactocentric distance. The bin with only one CC is not plotted because of the large error. From Fig. \ref{3.1.2.fig}, the most significant feature is the large difference between the rotation curves of the northern warp (red dots) and southern warps (blue dots). The northern warp has a lower rotational velocity and the southern warp has a higher rotational velocity compared to the mean rotational velocity at Galactocentric distances of $9-13$ kpc. The difference is also found in the radial velocities, which are shown in Fig. \ref{3.1.3.fig}. The northern warp has positive radial velocities while the southern warp has negative radial velocities. 

These velocity asymmetries may be due to warp precession, or to asymmetry in the structure of the northern and southern warps. Some previous works have found the north-south asymmetry around a Galactocentric distances of $8$ kpc that may be related to the resonance of the bar, the perturbation of the spiral arms \citep{2016MNRAS.457.2569M}, and the interaction with the Sagittarius dwarf galaxy or the LMC \citep{2013MNRAS.429..159G}. Based on the spiral arm potential of the perturbation distribution function, \cite{2016MNRAS.457.2569M} modelled the difference between the vertical and radial velocities of the stars on and between the spiral arms in the Galactocentric distances of $7-9$ kpc. Based on tidal interactions in Sagittarius dwarf galaxy, \cite{2013MNRAS.429..159G} modelled the asymmetry in the vertical direction locally at 8 kpc. The asymmetry in rotational and radial velocities that we found was mainly in the outer disk, which is more likely related to the warp. 

In contrast, the asymmetry in $V_{Z}$ is less obvious. In the warp model, the absolute value of the vertical velocity of CC is largest along the LONs and close to zero in directions with azimuthal angles of $\pm 90^{\circ}$. The projection map of $V_{Z}$ agrees well with the warp model, and $V_{Z}$ is around zero in the northern and southern warps.

\subsection{Cepheid kinematic warp model}
\label{sec:3.3}

Earlier in the paper, we mentioned that the shape of the Galactic disk in the vertical direction is warped, and that CCs are considered to be ideal probes for uncovering the warp formation and evolution \citep{2019NatAs...3..320C, 2019AcA....69..305S, 2023MNRAS.523.1556D}. In this section, we further investigate the Cepheid warp from the kinematic information.

Models of Galactic warp have been studied extensively in a number of works based on different tracers such as pulsars \citep{2004mim..proc..165Y}, red clump stars \citep{2002A&A...394..883L}, red giants \citep{2006A&A...451..515M}, etc. In general, the geometric model of warp can be expressed in the Galactocentric cylindrical coordinates ($R, \phi, Z$) as: 
\begin{linenomath*}
   \begin{equation} \label{eq7}
   \begin{aligned}
     Z_{w} = a(R-R_{w})^{b}\sin(\phi-\phi_{w}) \\
   \end{aligned}
   \end{equation}
\end{linenomath*}
where $Z_{w}$ is the average height of the stars above the plane of the Galactic disk, and $R_{w}$ and $\phi_{w}$ are the onset radius of warp and the Galactic azimuth of the warp's LONs, respectively. The parameters a, b are the fitting coefficients, representing the warp amplitude and the power index of the increase of the warp amplitude with the Galactocentric distance, respectively. 

We describe the warp precession model by introducing the evolution of the Galactic azimuth of the LONs $\phi_{w}(t)=\phi_{0,w}+\omega t$, where $\phi_{w}=\phi_{0,w}$ is the current LONs of the warp. This warp model is the simplest model for warp precession given by \cite{2020NatAs...4..590P}, who derived the expression of $\overline{V_{Z}}$ based on the zeroth moment of the collisionless Boltzmann equation. Ignoring the variation of the warp amplitude with time and assuming that $\omega$ does not depend on $R$, the expression for the warp velocity is: 
\begin{linenomath*}
   \begin{equation} \label{eq8}
   \begin{aligned}
     \overline{V_{Z}}(R, \phi) = \left( \frac{\overline{V_{\phi}}}{R}-\omega \right)\ a(R-R_{w})^{b}\cos(\phi-\phi_{w}) \\
   \end{aligned}
   \end{equation}
\end{linenomath*}
In this formula, $\overline{V_{\phi}}$ is the mean azimuthal velocity and $\omega$ denotes the precession rate for the Galactic azimuth of the LONs.

To discuss the warp model, we first fixed the Galactocentric distance to eliminate the variation of the warp amplitude with the Galactocentric distance. Here we chose the CCs with Galactocentric distances in the range of $12.5-13.5$ kpc, a total of 134 for further study. We calculated their average azimuth velocity $\overline{V_{\phi}}\approx236.81\ {\rm km\ s^{-1}}$ and put it into Equation (\ref{eq8}) together with $R = 13$ kpc. The parameters of the warp geometry model are directly taken from the Cepheid warp \citep{2019NatAs...3..320C} with $R_{w} = 9.26$, $\phi_{w} = 17.4$, $a = 0.148$ and $b = 1$. In this case, $\omega$ is the only free parameter in the kinematic warp model. 
At the same time, \cite{2019NatAs...3..320C} shows that there is a clear correlation between $a$, $b$ and $R_{w}$, but $\phi_{w}$ is weakly correlated with these three parameters. This suggests that the values of $a$, $b$ and $R_{w}$ for the warp model obtained from different studies may be different, while the parameter $\phi_{w}$ should be more consistent, i.e., $\phi_{w}$ is an independent parameter of the geometric warp model. From the kinematic warp model, it is known that stars on the warp's LONs have the maximum (minimum) vertical velocity. Therefore, we try to constrain the warp's LONs parameter $\phi_{w}$ and the warp progression rate $\omega$ by the vertical velocities of the CCs.

\begin{figure*}
\centering
  \begin{minipage}{185mm}
  \includegraphics[width=185mm]{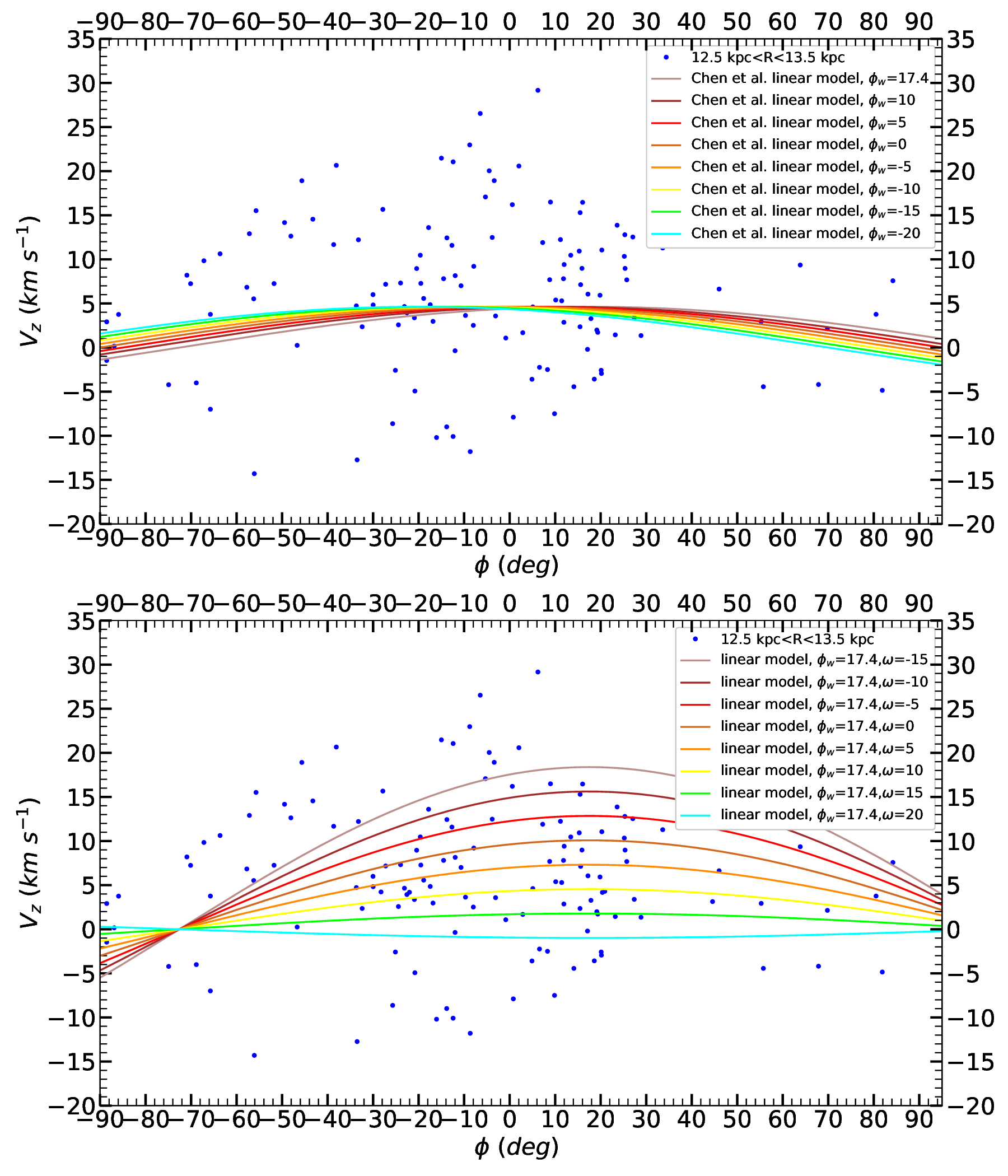}
\caption{Effects of warp's LONs $\phi_{w}$ and precession rate $\omega$ on the kinematic warp model. The blue dots show the distribution of 134 CCs with Galactocentric distances of $12.5-13.5$ kpc. The adopted warp geometric model is from \cite{2019NatAs...3..320C}. In the upper panel, the warp precession rate is fixed at $\omega=9.86$ km s$^{-1}$ kpc$^{-1}$ \citep{2020NatAs...4..590P}, with different color curves representing different $\phi_{w}$. In the lower panel, warp's LONs $\phi_{w}$ is fixed to $17.4^{\circ}$ and the different color curves represent different $\omega$.\label{3.3.1.fig}}
  \end{minipage}
\end{figure*}

We investigated the effects of $\phi_{w}$ and $\omega$ on the kinematic warp model in Fig. \ref{3.3.1.fig}. The upper panel of Fig. \ref{3.3.1.fig} shows the curves corresponding to different warp's LONs in $V_{Z}$ vs. $\phi$ space. We fixed the precession rate as $\omega = 9.86$ km s$^{-1}$ kpc$^{-1}$, which is from \cite{2020NatAs...4..590P} based on the warp linear model provided by \cite{2019NatAs...3..320C}. Our geometric parameters were taken from the linear model of \cite{2019NatAs...3..320C}. Unfortunately, the kinematic warp model is insensitive to the warp's LONs as seen in the figure, which means that the observed velocity does not constrain the warp's LONs well. Therefore, it's better to adopt the LONs from the geometric warp. In the lower panel of Fig. \ref{3.3.1.fig}, we fixed the warp's LONs to $17.4^{\circ}$ and used different $\omega$ to generate kinematic warp models. Here, the positive $\omega$ indicates the direction coincides with the rotation direction of the Milky Way. From the bottom panel of Fig. \ref{3.3.1.fig}, we can see that the curves of the kinematic warp models vary significantly with the precession rate, with higher precession rates corresponding to smaller vertical velocities. Therefore, the kinematic data of the CCs can well constrain the precession rate of the warp.

\begin{figure*}
\centering
  \begin{minipage}{180mm}
  \includegraphics[width=180mm]{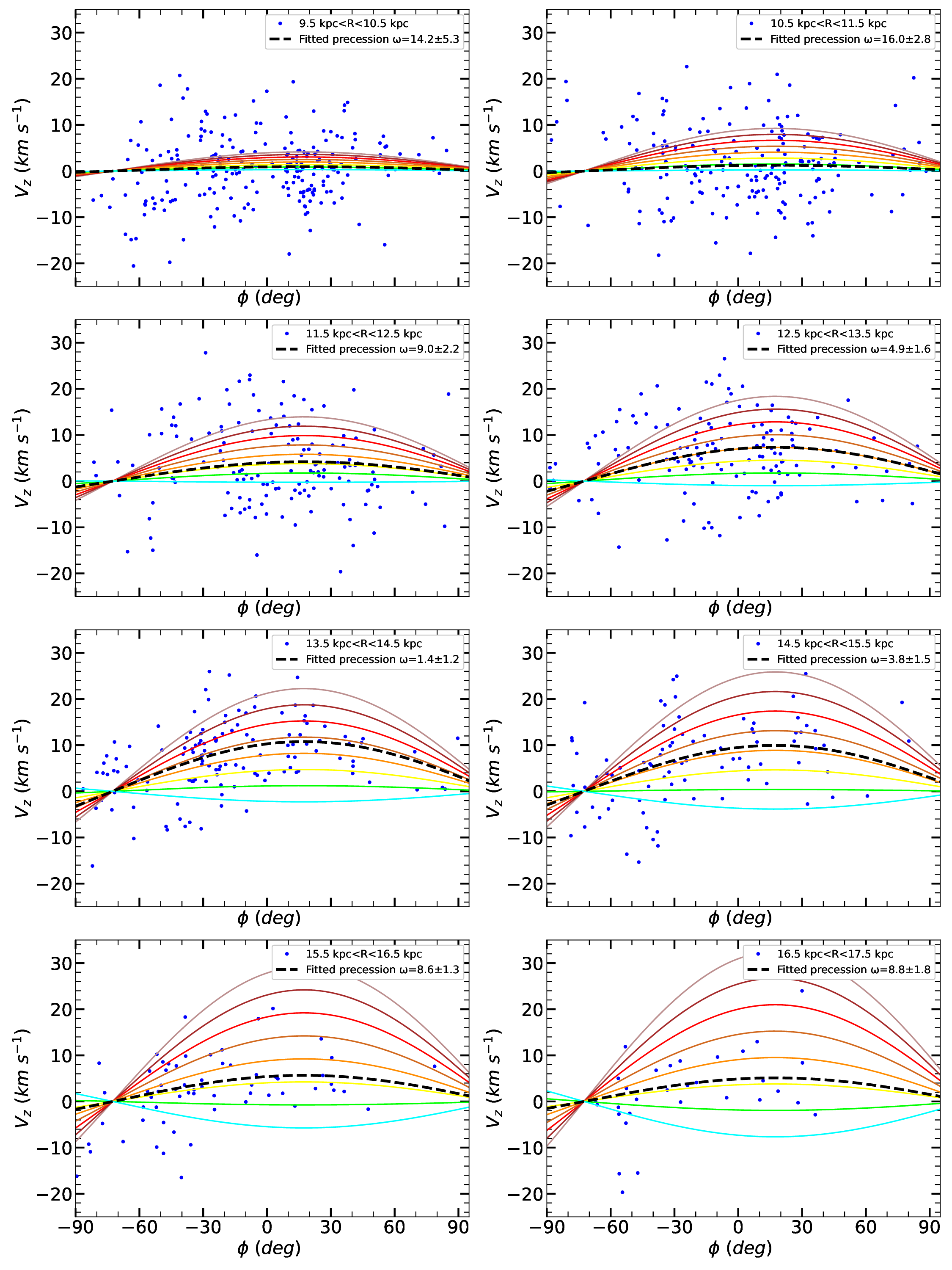}
\caption{Warp precession rates at different Galactocentric distances. The black dashed line is the best-fit warp precession rate and the results are given in the legend. The different colour curves are similar to the lower panel of Fig. \ref{3.3.1.fig} but for different Galactocentric distances of the warp model.\label{3.3.3.fig}}
  \end{minipage}
\end{figure*}

Subsequently, we further relaxed the radius parameter of the kinematic warp model to study the differences in the warp model at different Galactocentric distances. In Fig. \ref{3.3.3.fig}, we plot the variation of the kinematic warp model with precession rate at different Galactocentric distances. The parameters of the coloured solid lines in Fig. \ref{3.3.3.fig} are the same as in the lower panel of Fig. \ref{3.3.1.fig}, but for different Galactocentric distances. Since the kinematics of the CCs can well constrain the precession rate of the warp, we determined the optimal warp precession rates at each radius by least-squares fitting, as shown by the black dashed line in each panel of Fig. \ref{3.3.3.fig}. In the fitting process, the azimuth angle of the warp's LONs was set to $17.4^{\circ}$ and the other geometric parameters were consistent with the linear model of \cite{2019NatAs...3..320C}. 

The first three subplots in Fig. \ref{3.3.3.fig} show that the kinematic warp model is insensitive to changes in warp precession rate in the $9-12$ kpc ranges. The errors are too large to obtain an accurate warp precession rate. The kinematic warp model becomes sensitive to the warp precession rate as the Galactocentric distance increases, and the response is especially pronounced at azimuth near warp's LONs. However, from the two subplots at the bottom of Fig. \ref{3.3.3.fig}, we find that the number of CCs beyond 15 kpc is small, and the error due to completeness may dominate. Therefore, we propose to use the results of the Galactic warp precession rate at $12-14$ kpc.

\begin{figure*}
\centering
\begin{minipage}{185mm}
  \includegraphics[width=185mm]{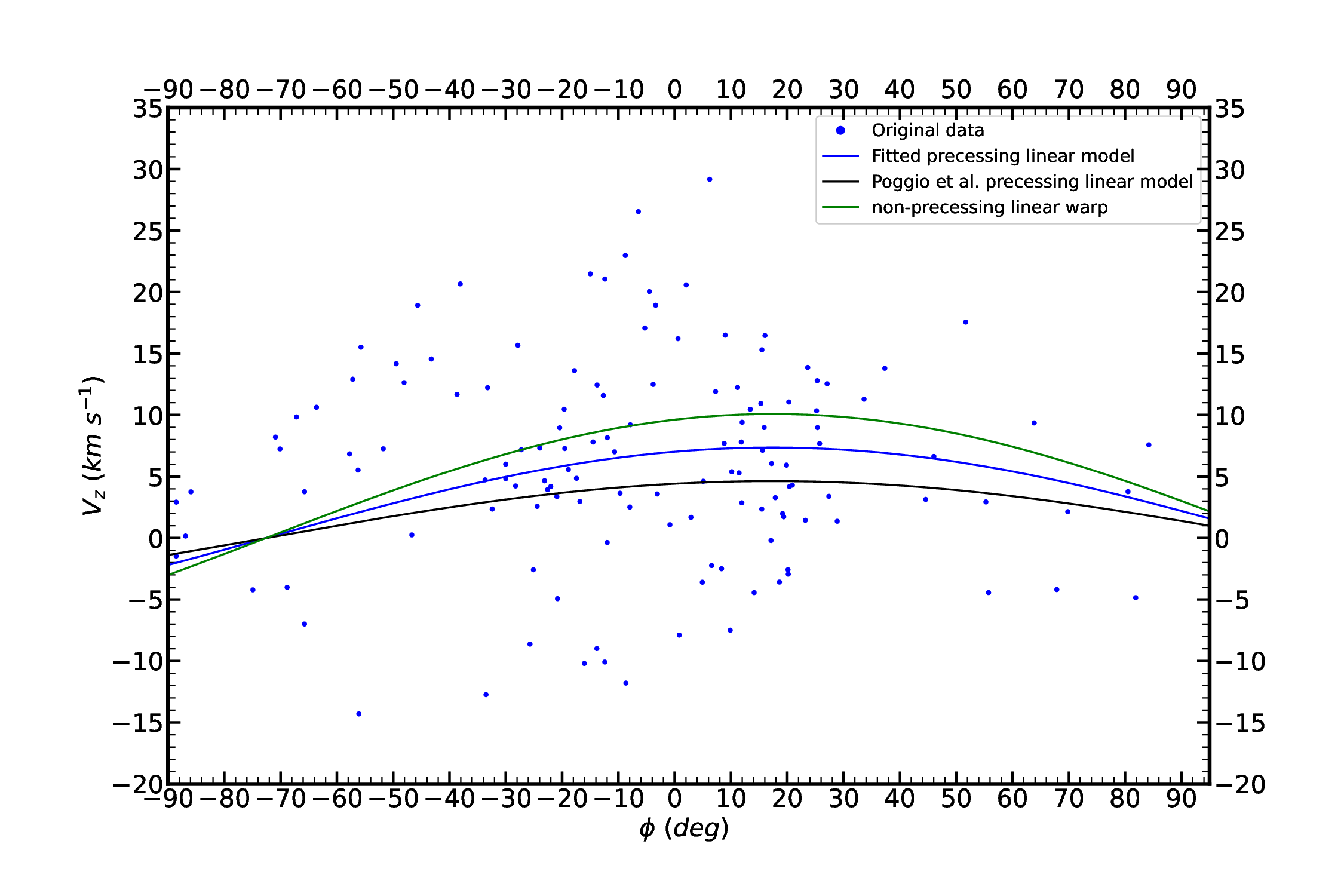}
\caption{Determination of the precession rate based on CCs. The blue line represents the best kinematic warp model obtained by fitting the kinematics of the CCs (blue dots) with the least squares method. The black line is the warp model with a higher precession rate $\omega = 9.86$ km s$^{-1}$ kpc$^{-1}$ given by \cite{2020NatAs...4..590P}, and the green line is the static warp model with $\omega = 0$ km s$^{-1}$ kpc$^{-1}$. 
 \label{3.3.2.fig}}
  \end{minipage}
\end{figure*}

Since \cite{2020NatAs...4..590P} used the sample at 13 kpc to obtain the precession rate, we finally chose to take the warp precession rate $\omega = 4.9\pm1.6$ km s$^{-1}$ kpc$^{-1}$ at 13 kpc as our result for the sake of comparison. We have indicated the best-fit result at 13 kpc in Fig. \ref{3.3.2.fig}, with a blue line, and using the black and green lines to indicate the warp precession model from \cite{2020NatAs...4..590P} and the static warp model, respectively. Our warp precession rate is lower than the result of \cite{2020NatAs...4..590P}. The Galactic warp precession rate given by \cite{2020NatAs...4..590P} is 10.86 km s$^{-1}$ kpc$^{-1}$, which is about 3.7 $\sigma$ larger than our result.  Based on the linear warp model of \cite{2019NatAs...3..320C}, \cite{2020NatAs...4..590P} gives a warp precession rate of 9.86 km s$^{-1}$ kpc$^{-1}$, which is 3.1 $\sigma$ larger than our results. Our results also exclude the static warp model with $\omega=0$, since the difference is 3.1 $\sigma$. 

In contrast, our precession rate agrees with the low warp precession rate of $\omega = 4$ km s$^{-1}$ kpc$^{-1}$ estimated by \cite{2021ApJ...912..130C} with a difference of only 0.6 $\sigma$. They used the whole sample of Gaia DR2, whose average age is much older than the CCs. \cite{2021ApJ...912..130C} suggested that \cite{2020NatAs...4..590P} determined their warp precession based on young population structure and old population kinematic information, leading to a systematic uncertainty in the precession rate. They found that younger populations have a larger warp amplitude and a higher vertical velocity, while older populations have the opposite. Therefore, using a combination of the low velocities of the older population and the large amplitude of the younger population yields a higher precession rate. In our study, we used both the geometric warp model and the kinematic warp model of CCs to determine the warp precession rate and our results support a low warp precession rate. Currently, the number of Galactic CCs with velocity information has greatly increased, providing us with a new opportunity to independently measure the Galactic warp precession rate based on CCs. In the future, based on more accurate and larger samples, the warp precession rate can be measured more accurately.

\section{Conclusions} \label{sec:concl}
Based on the RV information provided by Gaia DR3, we investigate the Galactic disk structure through CCs. The samples were taken from the Galactic CCs catalogues provided by \cite{2023A&A...674A..37G}. We extracted the samples with RV information from them and cross-matched them with the Gaia archive to obtain their astrometric and photometry data, with an initial sample of 2057 CCs. We estimated photometric distances for the CCs based on the period-Wessenheit-metallicity relation \citep{2022A&A...659A.167R}, and calculated the 3D positions and 3D velocities of CCs in the Galactic cylindrical coordinate system.

From the variation of the 3D velocities with the Galactocentric distance, we find the warp and flare feature from the vertical velocities, a peak shape in the radial velocities at $10-14$ kpc, and local fluctuations in the rotation curve. In the projection of the 3D velocity on the $XY$ plane, we clearly see the north-south asymmetry of the Galactic disk. To further investigate this phenomenon, we plot the rotational and radial velocities versus Galactocentric distances for CCs in different azimuthal ranges. The results show that the southern warp and the northern warp have significantly different patterns of rotational and radial velocity. These differences appear in the outer disk, and we propose that they may be related to the warp precession or the asymmetry of the warp structure.

We investigate the kinematic warp model by CCs and find that the kinematic warp model is insensitive to the parameters of the LONs, but sensitive to changes in the warp precession rate. The warp's LONs can be determined by the geometric information, while the vertical velocity can be used to constrain the warp precession rate. Meanwhile, we find that the kinematic warp model is insensitive to the warp precession rate up to 12 kpc, and as the distance increases the kinematic warp model becomes sensitive and the model response is largest near LONs. However, the current sample size beyond 15 kpc is too small, so we propose to use samples at $12-14$ kpc to obtain Galactic warp precession rates. Based on the warp linear model parameters of \cite{2019NatAs...3..320C}, we determine the warp precession rate at different Galactocentric distances. For comparison with the literature, we use the result at 13 kpc, $\omega = 4.9\pm1.6$ km s$^{-1}$ kpc$^{-1}$, as the final Galactic warp precession rate. Both the \cite{2020NatAs...4..590P}'s precession rate $\omega = 9.86$ km s$^{-1}$ kpc$^{-1}$ and the static warp model $\omega=0$ are outside the 3 $\sigma$ error range of our results, and we support a low warp precession rate as \cite{2021ApJ...912..130C}. Future Gaia data releases will provide a larger sample of CCs with RV curves. This will help to constrain the precession of the warp, thus providing more evidence to constrain the mechanisms of Galactic disk formation and evolution.

\section*{Acknowledgements}
This work was supported by the National Natural Science Foundation of China (NSFC) through grants 12173047, 11903045, 12003046, 12233009, 12133002, and 11973001. X. Chen and S. Wang acknowledge support from the Youth Innovation Promotion Association of the Chinese Academy of Sciences (no. 2022055 and 2023065). We also thank the support from the National Key Research and development Program of China, grants 2022YFF0503404 and 2019YFA0405504. This work presents results from the European Space Agency (ESA) space mission Gaia. Gaia data are being processed by the Gaia Data Processing and Analysis Consortium (DPAC). Funding for the DPAC is provided by national institutions, in particular the institutions participating in the Gaia MultiLateral Agreement (MLA). The Gaia mission website is \url{https://www.cosmos.esa.int/gaia}. The Gaia archive website is \url{https://archives.esac.esa.int/gaia}. 

\renewcommand{\bibname}{References}
\bibliographystyle{aasjournal}
\bibliography{ref}

\end{document}